\documentclass[iop]{emulateapj}
\usepackage{url}
\newcommand{\psr}{PSR~J1023$+$0038}
\newcommand{\first}{FIRST~J102347.67$+$003841.2}
\submitted{}

\begin{document}

\title{Evidence for gamma-ray emission from the low-mass X-ray binary system FIRST~J102347.6$+$003841}

\author{P.~H.~T. Tam$^{1}$, C.~Y. Hui$^{2}$, R.~H.~H. Huang$^{1}$, A.~K.~H. Kong$^{1,3}$, J. Takata$^{4}$, L.~C.~C. Lin$^{5}$, Y.~J. Yang$^{6}$, K.~S. Cheng$^{4}$, and R.~E.~Taam$^{7,8}$}
\affil {$^1$ Institute of Astronomy and Department of Physics, National Tsing Hua University, Hsinchu, Taiwan\\
$^2$ Department of Astronomy and Space Science, Chungnam National University, Daejeon, South Korea\\
$^3$ Kenda Foundation Golden Jade Fellow\\
$^4$ Department of Physics, University of Hong Kong, Pokfulam Road, Hong Kong\\
$^5$ Graduate Institute of Astronomy, National Central University, Jhongli, Taiwan\\
$^6$ Astronomical Institute ``Anton Pannekoek,'' University of Amsterdam, Amsterdam, The Netherlands\\
$^7$ Department of Physics and Astronomy, Northwestern University, 2131 Tech Drive, Evanston, IL 60208, U.S.A.\\
$^8$ Academia Sinica Institute of Astronomy and Astrophysics --- TIARA, Taipei, Taiwan\\}
\email{phtam@phys.nthu.edu.tw}

\begin{abstract}
The low-mass X-ray binary (LMXB) system FIRST~J102347.6$+$003841 hosts a newly born millisecond pulsar
(MSP) \psr\ that was revealed as the first and only known rotation-powered MSP in a
quiescent LMXB. While the system is shown to have an accretion disk before 2002, it remains unclear
how the accretion disk has been removed in order to reveal the radio pulsation in 2007.  In
this Letter, we report the discovery of $\gamma$-rays spatially consistent with FIRST~J102347.6$+$003841,
at a significance of 7 standard deviations, using data obtained by the Fermi Gamma-ray  Space Telescope.
The $\gamma$-ray spectrum can be described by a power law (PL) with a photon index of 2.9$\pm$0.2, resulting
in an energy flux above 200~MeV of $(5.5\pm0.9)\times10^{-12}\mathrm{erg}\,\mathrm{cm}^{-2}\,
\mathrm{s}^{-1}$. The $\gamma$-rays likely originate from the MSP \psr, but also possibly from an
intrabinary shock between the pulsar and its companion star. To complement the $\gamma$-ray study,
we also re-investigate the XMM-Newton data taken in 2004 and 2008. Our X-ray spectral analysis
suggests that a broken PL with two distinct photon indices describes the X-ray data significantly
better than a single PL. This indicates that there exists two components and
that both components appear to vary with the orbital phase. The evidence for $\gamma$-ray emission conforms
with a recent suggestion that $\gamma$-rays from \psr\ may be responsible for ejecting the disk
material out of the system.
\end{abstract}

\keywords{gamma rays: stars
                 --- Pulsars: individual (PSR~1023$+$0038)
                 --- X-rays: binaries}

\section{Introduction}

Radio millisecond pulsars (MSPs) are rotating neutron stars (NS) that have been spun up via the
transfer of angular momentum through accretion in low-mass X-ray binary (LMXB)
systems~\citep{Bhattacharya_review91}.
Recently, a total of 9 $\gamma$-ray emitting MSPs have been identified through their $\gamma$-ray pulsations~\citep{lat_millisecond_Science,lat_PSR0034}.

Theoretical models have long suggested that an accretion-powered MSP in a LMXB will turn on
as a rotation-powered MSP when the system is in quiescence~\citep{Alpar82}. How exactly the transition
happens is not completely understood, however, it
is widely believed that radio MSPs can only turn on after the accretion disk is removed. Suggested
mechanisms include the pulsar wind ablation~\citep{Wang09}, heating associated with the
deposition of $\gamma$-rays from the MSP~\citep{Takata_disk_10}, and the propeller effect~\citep{Romanova09}.

PSR~J1023$+$0038 is the first and only known rotation-powered MSP in a quiescent LMXB, namely \first\ (hereafter J1023). J1023 was identified as a LMXB in 2006~\citep{Homer06} and the radio MSP was found subsequently~\citep{1023_radiopsr_Science}. The source clearly showed an accretion disk before 2002~\citep{Wang09} and the disk has since disappeared~\citep{1023_radiopsr_Science}; radio pulsation was found in 2007~\citep{1023_radiopsr_Science}.
Therefore \psr\  is considered as a newly born MSP, representing the long sought-after missing link of a rotation-powered MSP descended from a LMXB.

The spin down power of \psr\ \citep[$L_{sd} \le 3 \times 10^{35}$~erg~s$^{-1}$;][]{1023_radiopsr_Science} is relatively high compared to other $\gamma$-ray MSPs~\citep{lat_millisecond_Science}. A small fraction of this power would suffice to generate detectable $\gamma$-rays from \psr. Given the recent evidence of an accretion disk before 2002 that later disappeared, it is of great importance to probe the energy source that facilitated the dissolution of the disk. The detection of X-rays from the system~\citep{Homer06} provides a hint that high-energy processes are ongoing, which is further strengthened by the reported X-ray pulsation from \psr\ and orbital variability in X-rays~\citep{1023_xray_pulse}. These results prompted us to search for $\gamma$-rays from the system and to study its X-ray properties in more detail.

\section{Gamma-ray observations and analysis results}

The Large Area Telescope (LAT) aboard the Fermi Gamma-ray Space Telescope is able to detect $\gamma$-rays with energies between $\sim$20~MeV and $>$300~GeV~\citep{lat_technical}. Data used in this work were obtained between 2008 August 4 and 2010 July 14 that are available at the Fermi Science Support Center\footnote{\url{http://fermi.gsfc.nasa.gov/ssc/}}. We used the Fermi Science Tools v9r15p2 package to reduce and analyze the data in the vicinity of J1023. Only those data that passed the most stringent photon selection criteria (i.e. the ``diffuse'' class) were used. To reduce the contamination from Earth albedo $\gamma$-rays, we excluded events with zenith angles greater than 105$^\circ$. The instrument response functions ``P6\_V3\_DIFFUSE'' recommended for analysis of the ``diffuse'' class events were used. We chose 200~MeV as the lower energy cut to include sufficient source photons at low energies while reducing the contamination of background photons that dominates at low energies. Therefore, we used events with energies between 200~MeV and 20~GeV in the likelihood analysis.

We carried out an unbinned maximum-likelihood spectral analysis (\emph{gtlike}) of the circular region of 15$^\circ$ radius centered on the $\gamma$-ray position (see below). We subtracted the background contribution by including the Galactic diffuse model (gll\_iem\_v02.fit) and the isotropic background (isotropic\_iem\_v02.txt), as well as all sources in the first Fermi/LAT catalog~\citep[1FGL;][]{lat_1st_cat} within the circular region of 25$^\circ$ radius around the $\gamma$-ray position. We assumed a power law (PL) spectrum for all the 52 1FGL sources considered. The spectral parameter values were set free for sources within 10$^\circ$ from J1023.

\begin{figure}
    \epsscale{1.}
    \plotone{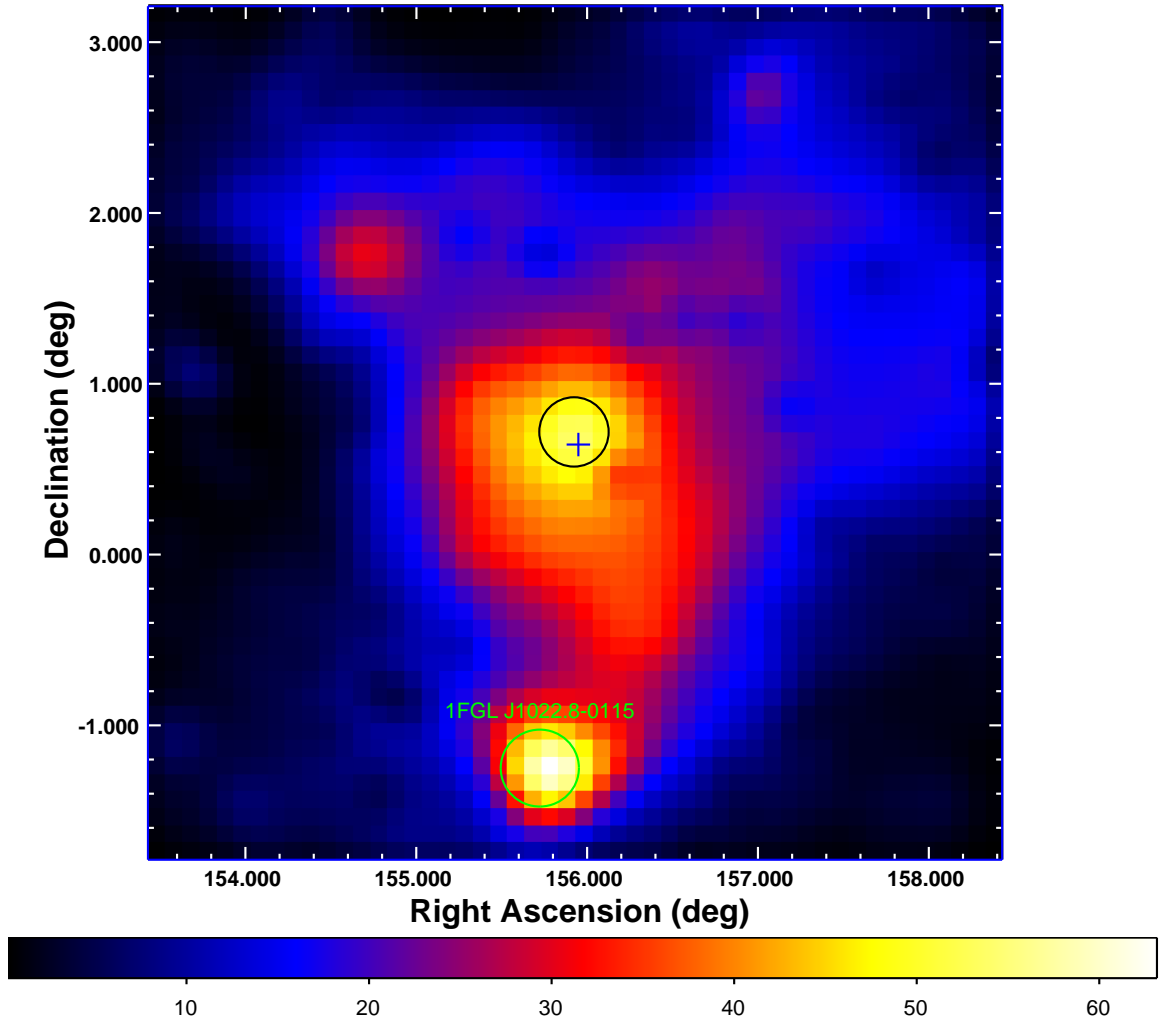}
      \caption{Test-statistic (TS) map of a region of 5$^\circ\times$5$^\circ$ centered at the position of J1023 (labeled by the cross). Gamma-rays with energies between 200~MeV and 20~GeV were used. This map is created by moving a putative point source through a grid of locations on the sky and maximizing $-$log(likelihood) at each grid point, while stronger and well-identified sources outside the sky map are included in each fit. The Fermi source 1FGL~J1022.8-0115 is clearly seen in the map. The 95\% confidence-level error circles of the best-fit position of the $\gamma$-ray emission are also shown. The error circle of 1FGL~J1022.8-0115 is taken from~\citet{lat_1st_cat}.}
      \label{1023_TSmap}
   \end{figure}

The maximized \emph{test-statistic} (TS) value~\citep{Mattox_96} we obtained for the pulsar position is 50, corresponding to a detection significance of 7$\sigma$. The position of the $\gamma$-ray source is estimated by \emph{gtfindsrc} to be at R.A. (J2000) $=$ 155$\fdg$92 and Dec. (J2000) $=$ 0$\fdg$72 with statistical uncertainty 0$\fdg$08 (0$\fdg$2) at the 68\%(95\%) confidence level, which is consistent with the position of J1023. The systematic uncertainty is estimated to be $\leq$40\%~\citep{bsl_lat}. We used \emph{gttsmap} to obtain the \emph{TS map} of the 5$^\circ\times$5$^\circ$ region centered on the best-fit $\gamma$-ray position, as shown in Fig.~\ref{1023_TSmap}.

To investigate why this $\gamma$-ray source was not present in the 1FGL catalog, we divided the whole dataset into two: the first and second year data, respectively, and performed the likelihood analysis for each of them separately. Both TS values drop to 22--24, which are about half of the TS value derived using the whole dataset and below the threshold of TS$=$25 to be included in the 1FGL catalog~\citep{lat_1st_cat}. This suggests that the source we found is significant (i.e., $>$5$\sigma$) with two-year data. The spectral parameters from each dataset are also consistent with the values reported below.

We then fit the $\gamma$-ray spectrum with a single PL, resulting in a photon index of $2.9\pm0.2$ and an integrated energy flux above 200~MeV of $(5.5\pm0.9)\times10^{-12}$~erg~cm$^{-2}$~s$^{-1}$. We divided the 200~MeV--20~GeV $\gamma$-rays into five energy bins of logarithmically equal bandwidths and reconstructed the flux using \emph{gtlike} for each band separately, assuming a PL model with $\Gamma_\gamma=$2.9 within each bin. No $\gamma$-ray source was needed at the J1023 position in the three bins above 1.3~GeV (derived TS values $<$5) in the likelihood analysis, indicating a cut-off at $\sim$1~GeV. We therefore attempted to fit the 200~MeV--20~GeV spectrum with a PL with an exponential cut-off (PLE). We found that PLE also well describes the spectrum, i.e. both PL and PLE models gave the same TS value of 50. However, it should be noted that a spectral cutoff is statistically not required. Due to the low photon statistics ($\sim$280 modeled photons from J1023), the spectral parameters of the PLE model cannot be well constrained simultaneously. Motivated by the possible magnetospheric origin of the $\gamma$-rays, $\Gamma_\gamma$ and $E_\mathrm{c}$ are estimated by fixing the other one (at its mean value) to be 1.9$\pm$0.3 and 700$\pm$220~MeV. We fixed $\Gamma_\gamma$ at 1.5 to 2.4 (with steps of 0.1) while letting the normalization and $E_\mathrm{c}$ free, and found that $E_\mathrm{c}$ was only well constrained when $\Gamma_\gamma=$~1.7 to 2.1, consistent with the above result.

At the distance of 1.3~kpc, the $\gamma$-ray luminosity (above 200 MeV) is $(1.1\pm0.2)\times10^{33}$~erg~s$^{-1}$. Assuming that the $\gamma$-rays come from \psr\ (see Sect.~\ref{sect:discussion}), the pulsar spin-down luminosity $\dot{E}<3\times10^{35}$~ergs~s$^{-1}$~\citep{1023_radiopsr_Science} implies a $\gamma$-ray conversion efficiency of only $\ga$0.3\%. Such $\gamma$-ray luminosity and $\gamma$-ray efficiency are among the smallest of $\gamma$-ray MSPs. Moreover, assuming the PLE model is robust, the cutoff energy of $\sim$700~MeV is the lowest among all $\gamma$-ray MSPs. The spectral properties of the $\gamma$-ray emission from J1023 are summarized in Table~\ref{gamma_spec_par}.

A search for pulsation of the $>$10~MeV $\gamma$-rays within a 1$\degr$~radius region around the $\gamma$-ray position was performed. We did not find any significant pulsed detection at or close to the spin-period of the pulsar nor any indication of $\gamma$-ray variability related to the orbital modulation. Even so, given the low photon statistics (of just over 700 photons), this result does not preclude any $\gamma$-ray pulsation.

We also performed a long-term temporal analysis of J1023, in which the 2-year data were binned in 3-month periods. No significant $\gamma$-ray variability was found, indicating that the object is stable (down to 3-month periods) in radiating $\gamma$-rays.

\section{A model of $\gamma$-ray emission from \psr}
\label{sect:discussion}
The $\gamma$-rays from J1023 originate either from the pulsar magnetosphere or a shock where material overflowing from the companion interacts with the pulsar wind. In the pulsar wind scenario, $\gamma$-rays may be generated as synchrotron radiation of the electrons and positrons accelerated in the shock~\citep{Takata09}. An important theoretical uncertainty is how a particle kinetic-dominated flow can be formed near the pulsar~\citep{Kirk03,Arons08}. In addition, it is also difficult to explain the steep PL spectrum ($\Gamma_\gamma\sim$3) in the synchrotron model. On the other hand, if the pulsar magnetosphere is sufficiently clear of matter, GeV $\gamma$-ray photons can be produced in a slot-gap~\citep{Harding05} or an outer-gap accelerator~\citep{Cheng86a,Cheng86b}. The curvature radiation from a gap accelerator is expected to have a PLE shape. The reported X-ray pulses from \psr\ \citep{1023_xray_pulse} suggests that the observed $\gamma$-rays are also produced in the pulsar magnetosphere.
More accumulated source photons in the future should help to answer whether there exists $\gamma$-ray pulsation or variability on the time scale of the orbital period, in turn helping to distinguish these two scenarios for the observed $\gamma$-rays.

Applying the outer gap model, the $\gamma$-ray luminosity is approximately described by
\begin{equation}
L_{\gamma}\sim f^3 L_{sd},
\end{equation}
 where $f$ is the fractional gap thickness, i.e. the ratio of the gap
thickness to the radius of the light cylinder and $L_{sd}=4(2\pi)^4B^2R^6/6c^3P^4$ the pulsar spin down power
 with $B$ corresponding to the surface magnetic field, $R$ the stellar radius, and $P$ the rotation period.  The fractional gap thickness is determined by the pair-creation condition between $\gamma$-rays emitted in the outer gap and X-rays from the NS surface,
 i.e., $E_{\gamma}E_{s,X}=(m_ec^2)^2$, where $E_{\gamma}\sim 1$~GeV and $E_{s,X}$ are the typical energy of the
$\gamma$-rays and the X-rays, respectively.  Assuming $E_{s,X}=30$~eV (see Sect.~\ref{sect:xray} and
Table~\ref{x_spec_par}), and following~\citet{Takata10} we find f $\sim 0.5$.
The $\gamma$-ray luminosity is estimated as $L_{\gamma}\sim 4\times 10^{34}$erg~s$^{-1}$ using $L_\mathrm{sd}=3\times 10^{35}~\mathrm{erg~s^{-1}}$, consistent with
the required luminosity from an irradiating source to explain the heating of the companion star~\citep{Thorstensen05}.

We may estimate the $\gamma$-ray flux measured on the Earth as $F_{\gamma}\sim L_{\gamma}/\delta\Omega
d^2\sim 10^{-9}~\mathrm{erg/cm^2 s}$, where $L_{\gamma}=3\times 10^{34}~\mathrm{erg~s^{-1}}$, the solid angle
$\delta\Omega=2$, and the distance $d=1.3$~kpc are assumed. This flux is higher than the observed $F_{\gamma}\sim 5\times 10^{-12}~\mathrm{erg~cm^{-2}~s^{-1}}$. We note, however, that the $\gamma$-ray flux
depends on the viewing geometry, because the intensity varies over the $\gamma$-ray beam.
We carried out a simple 3-D calculation using curvature radiation in the outer gap model~\citep[c.f.][]{Wang_Takata_Cheng10,Takata07} and found that $F_{\gamma}$ increases from $\sim$5$\times10^{-13}~\mathrm{erg~cm^{-2}~s^{-1}}$ to $\sim$10$^{-11}~\mathrm{erg~cm^{-2}~s^{-1}}$ when the viewing angle, i.e. the angle between the rotational axis of the pulsar and the line of sight, $\xi$, increases from $\sim$34$^{\circ}$ to $\sim$53$^{\circ}$ \citep[see][]{1023_radiopsr_Science}, well consistent with the observed $\gamma$-ray flux.

\section{Re-analysis of the X-ray data}
\label{sect:xray}
XMM-Newton observations of J1023 were conveyed on 12 May 2004 (hereafter
we refer this observation as XMM1) with all the EPIC cameras operated in full-frame mode, and on 26 November 2008 (hereafter XMM2) with the MOS~1/2 CCDs operated in full-frame mode and the PN camera operated
in timing mode with a resolution of 0.03~ms.

Archibald et al. (2010) reported an X-ray analysis of J1023, finding a
possible X-ray pulsation from \psr\ and modulation on the orbital period of the binary.  They
interpreted the latter as coming from an intrabinary shock, and suggested a composite spectral model with a PL plus
a possible thermal component. Although this model results in an
acceptable goodness-of-fit, we found that their analysis is apparently incomplete.
The X-ray spectrum from the pulsar magnetosphere is typically different from
the shock emission (cf. Hui \& Becker 2006, 2007, 2008), and the combined contribution from the two emission regions cannot be generally described by a single PL.
To complement their results, we report our independent
analysis of these XMM-Newton observations.

As PN data obtained from XMM2 are collapsed into a one-dimensional row, they are not suitable for spectroscopy. After filtering for the high sky background and events affected by bad pixels,
the effective exposures are 14.9~(MOS~1/2) and 11.6~kiloseconds (PN) for XMM1, and 33.5~kiloseconds (MOS1/2) for XMM2.

To avoid contamination from a nearby X-ray source, we extracted
the energy spectrum of J1023 from the circles with radius 25''
~\footnote{Circles of radius 35'' were used in~\citet{1023_xray_pulse}, that probably contain
a larger amount of contamination from the PSF wing of that nearby X-ray source.} centered on
the radio timing position in all datasets, which
corresponds to an encircled energy fraction $\sim80\%$. The background
spectra were sampled from nearby low-count circular regions
of radius 40'' in the corresponding cameras. 
We grouped each spectrum dynamically to obtain the same signal-to-noise ratio in
each dataset.

First, we fit the XMM1 and XMM2 data separately with various single-component models. Only the PL model provides
a good description of the data in both XMM1\footnote{Our results for XMM1 are consistent with those
reported by Homer et al. (2006).} and XMM2 (see Table~\ref{x_spec_par}).
With no indication of spectral and flux variability between data taken in XMM1 and XMM2, we
combined both datasets for a constraining spectral analysis.

\begin{figure}
\includegraphics[angle=-90,width=\columnwidth]{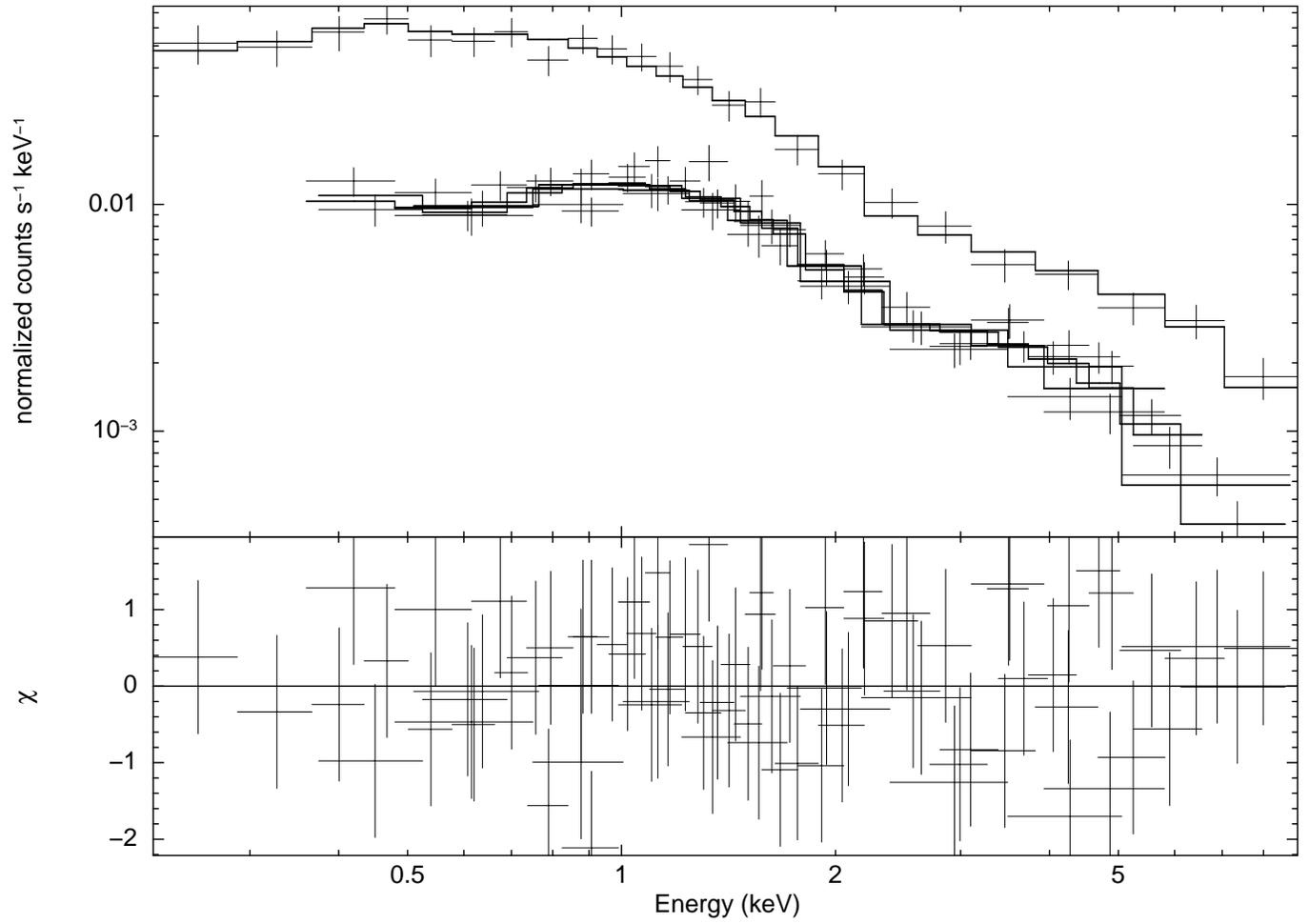}
\caption{The X-ray spectra of J1023 as obtained from the observations
taken at 12 May 2004 and 26 November 2008 with XMM-Newton and simultaneously fitted to an absorbed
broken power-law model (upper panel) and contribution to the $\chi^{2}$ fit statistic
(lower panel).}
\label{xray_spec}
\end{figure}

We found that the single PL model provides an acceptable description of the combined data with the
spectral parameters as in Archibald et al. (2010). However, we found systematic deviations in this model, indicating that additional component(s) might be required.
Motivated by the possible presence of both pulsed emission and emission dependent on the orbital phase of the binary, we fit the spectrum with
a broken power law (BKPL). The goodness-of-fit was found to be improved significantly ($\chi^{2}_{\nu}=0.75$ for 75 d.o.f.) and no systematic fitting residuals were found (see Fig.~\ref{xray_spec}), in contrast with the best-fit model (i.e. single PL model) presented in Fig. 1 of~\citet{1023_xray_pulse}. To statistically address the improvement for the spectral fits from the single PL to the broken PL, we used an F-test and found that the p-value is 7.7$\times10^{-6}$. Therefore, the additional parameters are required statistically.
This behavior was found in both epochs (XMM1 as well as XMM2). The break energy was found to be $E_{b}=1.84^{+0.22}_{-0.16}$~keV. The spectrum was found to be steeper
(i.e. $\Gamma^{1}_{X}=1.75^{+0.16}_{-0.11}$) at $E<E_{b}$ than that in the hard band (i.e.
$\Gamma^{2}_{X}=1.07^{+0.07}_{-0.06}$). The unabsorbed flux was found to be
$f_{X}=6.3^{+1.6}_{-0.9}\times10^{-13}$~ergs~cm$^{-2}$~s$^{-1}$ (0.01--10~keV).
At a distance of $d\sim1$~kpc, the isotropic X-ray luminosity is
$L_{X}=7.5^{+1.9}_{-1.1}\times10^{31}$~ergs~s$^{-1}$ (0.01--10~keV), corresponding to an X-ray
conversion efficiency of $\ga0.03\%$ of the spin-down luminosity.

As thermal X-rays from the NS surface may also contribute,
we estimated the thermal contribution besides the NMHA model
with the mass and radius of the NS fixed at $M_{\rm NS}=1.4~M_{\odot}$
and $R_{\rm NS}=10$~km. Nevertheless, this additional thermal model was only required at a confidence
level of $<50\%$. Furthermore, with normalization as a free parameter, an unreasonably small source distance of
$\sim30$~pc was inferred. To constrain the possible thermal contribution from the
stellar surface, we fixed the normalization corresponding to 1~kpc. This results in the $1\sigma$ upper limits
of the temperature and the thermal flux contribution to be $kT<32.76$~eV and
$f_{X}<7.3\times10^{-14}$~ergs~cm$^{-2}$~s$^{-1}$ (0.01--10~keV), respectively.

The fact that more than one PL index is required to explain the observed X-ray spectrum strongly suggests an additional X-ray contribution from this system besides the magnetospheric radiation. As possible orbital modulation in X-rays has been identified,
the additional non-thermal component may come from the intrabinary shock~\citep{1023_xray_pulse}.
Furthermore, our spectral analysis suggests that the emission below and above the break energy ($\sim$2~keV)
might have different origins. \citet{1023_xray_pulse} reported an X-ray pulsation for photons at 0.25--2.5~keV. Whether the emission above $\sim$2~keV is pulsed remains unclear.

\begin{figure}
\includegraphics[angle=-90,width=\columnwidth]{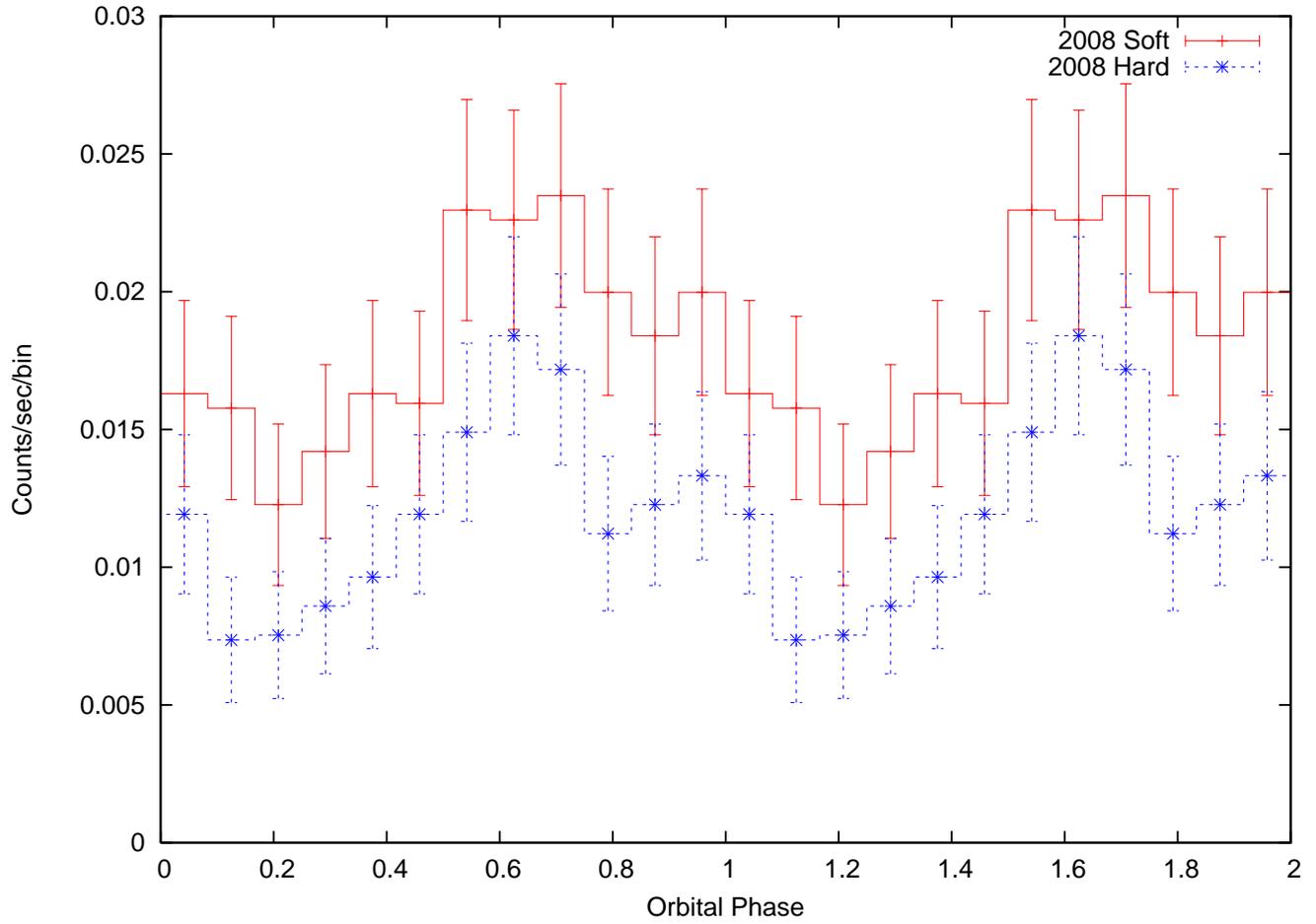}
\caption{The energy-resolved X-ray light curves of J1023 in the soft (0.3--2.0~keV) and hard (2.0--10.0~keV) bands, as obtained from the observation taken at 26 November 2008 with XMM-Newton.}
\label{xray_orbital}
\end{figure}

We performed an energy-resolved analysis of the
X-ray flux as a function of orbital phase. According to the resultant break energy, we divided the
energy range (0.3--10~keV) suitable for timing analysis into two bands: 0.3--2~keV (soft) and
2--10~keV (hard).
We found no significant variability at the orbital period in both bands from XMM1 data,
consistent with~\citet{Homer06}. This is likely due to low photon statistics, as well as insufficient orbital coverage of $\sim$91\% for MOS1/2 and $\sim$82\% for PN data. We therefore focused on the XMM2 data in this study. The orbital light curves of the soft and hard bands are shown in Fig.~\ref{xray_orbital}. Both soft and hard X-rays contribute to the orbital modulation and the troughs and peaks in the orbital light curves in these two bands occur at the same orbital phases (around 0.2 and 0.6 respectively). Using a
$\chi^{2}$-test, the significance for a flux modulation over the observed orbit in the soft and hard bands
was found to be $\sim$99.1\% and $\sim$85.5\%, respectively. Although the significance derived from this test is higher for soft X-rays than for hard X-rays, we could not determine the energy band for which the modulation
is greater.

\section{Concluding remarks}
We have found strong evidence for $\gamma$-ray emission from a LMXB system, FIRST~J102347.67$+$003841.2. The $\gamma$-rays may originate from \psr, or possibly from the intrabinary shock between the pulsar and its companion star. Given the observed steep spectrum ($\Gamma_\gamma\sim$3), synchrotron
emission from the shock is not favored. We have applied an outer gap model to explain the observed $\gamma$-rays. Recently, \citet{Takata_disk_10} suggest that such $\gamma$-rays from a newly-born MSP may be responsible for the disk clearance in such systems.
Our discovery of $\gamma$-rays, once confirmed, from FIRST~J102347.67$+$003841.2 conform with their predictions.

The spectral analysis of the XMM data taken in 2004 and 2008 suggests that a broken PL well describes the X-ray data, with the lowest systematics, compared to a single PL and other composite models tested. Our X-ray spectral analysis suggests that there exists two components and
that both components appear to vary with the orbital phase, supporting the X-ray variability study by~\citet{1023_xray_pulse}, that suggested more than one mechanism contributes to the observed X-ray flux from J1023.

In such a system that recently transited from the LMXB phase to the radio MSP phase, we speculate that the current
state may not be permanent as an accretion disk may reform about the NS in \psr\, switching off the radio MSP. A
cycle of on and off states of pulsar activity could repeat in the near future. Continuous monitoring in all wavelengths of this unique source will test this idea and help us to comprehend the on-going evolution of FIRST~J102347.67$+$003841.2.

\acknowledgments
We acknowledge the use of data and software facilities
from the FSSC, managed by the HEASARC at the Goddard Space Flight Center. CYH is supported by research
fund of Chungnam National University in 2010. AKHK and LCCL are supported partly by
the National Science Council of the Republic of China
(Taiwan) through grant NSC99-2112-M-007-004-MY3 and NSC99-2811-M-008-057, respectively. YJY has received funding from the European Community's Seventh Framework Programme (FP7/2007-2013) under grant agreement number ITN 215212 ``Black Hole Universe''. KSC is supported by a GRF grant of Hong Kong Government under HKU700908P.

\clearpage

\begin{deluxetable}{lccccc}
\tablecaption{Spectral parameter values of $\gamma$-ray emission from J1023. \label{gamma_spec_par}}
\tablewidth{0pt}
\tablehead{
\colhead{Model\tablenotemark{a} } & \colhead{ Photon flux\tablenotemark{b} ($>$200~MeV)} & \colhead{Energy flux ($>$200~MeV)} & \colhead{Photon Index} & \colhead{Cutoff energy} & \colhead{$\gamma$-ray luminosity\tablenotemark{c} ($>$200~MeV)} \\
                            & \colhead{($10^{-9}$~cm$^{-2}$~s$^{-1}$)} & \colhead{($10^{-12}$~erg~cm$^{-2}$~s$^{-1}$)} & \colhead{$\Gamma_\gamma$} & \colhead{(MeV)  } & \colhead{($10^{33}$~erg~s$^{-1}$)}
}
\startdata
    PL      & $8.2\pm1.6$ & $5.5\pm0.9$ & $2.9\pm0.2$ & --- & $1.11\pm0.18$ \\
\tableline
    PLE     & $8.1\pm1.5$ & $5.2\pm0.8$ & 1.9\tablenotemark{d} & $700\pm230$ & $1.06\pm0.17$ \\
    PLE     & $7.8\pm1.5$ & $5.1\pm0.8$ & $1.9\pm0.3$ & 700 & $1.03\pm0.17$ \\
\enddata
\tablenotetext{a}{~PL$=$power law model; PLE$=$power law with an exponential cut-off model}
\tablenotetext{b}{~All the quoted errors are statistical and $1\sigma$ for one parameter of interest.}
\tablenotetext{c}{~The pulsar distance is taken as 1.3~kpc.}
\tablenotetext{d}{~Model parameters without quoted errors are fixed at the value given.}
\end{deluxetable}


\begin{deluxetable}{l@{}ll@{}cl@{}c@{}l}
\tabletypesize{\scriptsize}
\tablecaption{Spectral parameter values of X-ray emission from J1023. \label{x_spec_par}}
\tablewidth{0pt}
\tablehead{
\colhead{Model\tablenotemark{a}} & \colhead{Epoch      } & \colhead{$n_\mathrm{H}$\tablenotemark{b}} & \colhead{Photon Index\tablenotemark{c}} & \colhead{$E_\mathrm{b}/kT$} & \colhead{Unabsorbed flux\tablenotemark{d}} & \colhead{$\chi^{2}_{\nu}$ (d.o.f.)} \\
                   &     & \colhead{(10$^{20}$~cm$^{-2}$)} & \colhead{$\Gamma_X$} & \colhead{(keV)} & \colhead{($10^{-13}$erg~cm$^{-2}$s$^{-1}$)} &
}
    \startdata
    PL  & 12 May 2004 (XMM1) & $<$1.2     & $1.30^{+0.04}_{-0.05}$ & $.../...$ & $4.6^{+0.5}_{-0.3}$ & 0.94 (40) \\
    PL  & 26 Nov 2008 (XMM2) & $<$0.4     & $1.29\pm0.04$  & $.../...$ & $4.9^{+0.4}_{-0.3}$    & 1.17 (35) \\
    \tableline
    PL  & XMM1 \& XMM2 & $<$0.6  & $1.29\pm0.03$  & $.../...$ & $4.7^{+0.3}_{-0.2}$    & 1.00 (77) \\
    BKPL  & XMM1 \& XMM2 & $4.4\pm1.6$  & $1.75^{+0.16}_{-0.11}(E<E_{b})/1.07^{+0.07}_{-0.06}(E>E_{b})$  & $1.84^{+0.22}_{-0.16}$/... & $6.3^{+1.6}_{-0.9}$    & 0.75 (75) \\
    BKPL  & XMM1 \& XMM2 & $4.7^{+3.3}_{-1.9}$ & $1.76^{+0.17}_{-0.14}(E<E_{b})/1.07^{+0.07}_{-0.08}(E>E_{b})$  & $1.84^{+0.14}_{-0.15}$/$<$0.033 & $6.4^{+1.8}_{-0.9}$ (BKPL) & 0.76 (74) \\
    \& NMHA                       &                       &             &             &    & $<0.73$ (NMHA) & \\
\enddata
\tablenotetext{a}{~PL$=$power law model; BKPL$=$broken power law model; NMHA$=$non-magnetic hydrogen atmospheric model}
\tablenotetext{b}{~column absorption}
\tablenotetext{c}{~All quoted errors are $1\sigma$ for one parameter of interest.}
\tablenotetext{d}{~in the 0.01--10~keV range}
\end{deluxetable}

\clearpage

\centering

\end{document}